\documentclass[twocolumn,showpacs,preprintnumbers,amsmath,amssymb]{revtex4}

\usepackage{graphicx}
\usepackage{epstopdf}
\usepackage{dcolumn}
\usepackage{bm}

\usepackage{amsmath}
\usepackage{amsfonts}
\usepackage{amssymb}

\begin{document}


\title{Cosmogenic Tau Neutrino Induced Radio Emission}

\author{Kwang-Chang Lai}
 \email{kclai@mail.nctu.edu.tw}
\affiliation{Institute of Physics, National Chiao Tung University,
Hsinchu, 300, Taiwan.\\
Leung Center for Cosmology and Particle Astrophysics, National Taiwan University,
1 Sec. 4, Roosevelt Rd., Taipei,106, Taiwan}

\author{Guey-Lin Lin}
 \email{glin@cc.nctu.edu.tw}
\affiliation{Institute of Physics, National Chiao Tung University,
Hsinchu, 300, Taiwan.\\
Leung Center for Cosmology and Particle Astrophysics, National Taiwan University,
1 Sec. 4, Roosevelt Rd., Taipei,106, Taiwan}

\author{Tsung-Che Liu}
 \email{diewanger@gmail.com}
\affiliation{Institute of Physics, National Chiao Tung University,
Hsinchu, 300, Taiwan.}

\author{JiWoo Nam}
 \email{jiwoonam@gmail.com}
\affiliation{Department of Physics, Ewha Womans University, Seoul, Korea}

\author{Chi-Chin Chen}
 \email{b88202054@ntu.edu.tw}
\affiliation{Institute of Astrophysics, National Taiwan University,
1 Sec. 4, Roosevelt Rd., Taipei,106, Taiwan}

\date{\today}

\begin{abstract}

Cosmogenic neutrinos\cite{beren} are expected from ultrahigh energy cosmic rays undergoing the GZK process\cite{GZK1,GZK2} and anticipated to be observed by detecting air showers from the decays of tau leptons. We use CORSIKA simulated shower structure to calculate the coherent geosynchrotron radio emissions of the tau decay showers above $10^{17}$eV. We present the pattern and spectrum of radio waves and discuss their detections by radio antennae.

\end{abstract}

\maketitle

\section{Introduction}

The origin of ultra-high energy cosmic rays remains a fundamental and unsolved problem in astroparticle physics. Promising clues could be provided by the associated high energy neutrinos since they would neither interact with intergalactic or interstellar media nor be deflected by the magnetic fields. Various detectors have been proposed for detecting high energy neutrinos. Some of them rely on measuring the air shower by the so-called earth-skimming $\nu_\tau$, for which horizontal showers are generated by the ensuing $\tau$ decay\cite{Domokos:1997ve,Fargion}. In this paper, we investigate the shower properties by simulations. Equipped with the knowledge of  the $\nu_\tau$ induced air shower, we are able to calculate the induced geosynchrotron radiation.

 In Sec. II, we present the CORSIKA\cite{cor} simulated shower profile to be employed in the calculation of geosynchrotron radiation in Sec. III. Our calculation is based on the coherent geosynchrotron emission scenario initiated in 1970's\cite{Allan} and further developed by Huege and Falcke\cite{HF}. In Sec. IV, we summarize and conclude our work.

\section{Air Shower Simulations}

The tau decay induced air shower is initiated by the decay product. Using  CORSIKA code, we simulate the  shower initiated by electrons at five different energies. Table \ref{stat} presents the statistics of these simulations. The simulation shows that the shower particles reside in a shower thickness less than $1{\rm m}$. Compared with the radiation which traverses a distance of $\sim10$ ${\rm km}$, the shower front at the shower maximum is treated as longitudinally coherent. The remaining structures are the lateral profile and Lorentz factor distribution representing the spatial and energy distribution of the shower particles.

\begin{table}[htb]
  \centering 
  \caption{Shower statistics }\label{stat}
  \begin{tabular}{lccc}\\
\hline
   Shower              &  \multicolumn{3}{c}{ number of $e^-$ and $e^+$}     \\  \cline{2-4}
   energy (eV)              &  Total                        &   $\gamma=1-100$    &     $\gamma=1-1000$  \\  \hline
   $10^{17}$                &  $7.32\times 10^7$                       &    $4.54\times 10^7$          &     $6.79\times 10^7$          \\ 
   $10^{17.5}$            &  $2.23\times 10^7$                       &    $1.34\times 10^8$          &     $1.99\times 10^8$          \\ 
   $10^{18}$               &  $7.10\times 10^8$                       &    $4.41\times 10^8$          &     $6.55\times 10^8$          \\ 
   $10^{18.5}$            &  $2.15\times 10^8$                       &    $1.32\times 10^9$          &     $1.98\times 10^9$          \\ 
\hline
\end{tabular}
\end{table}

\begin{figure}[htb]
\begin{center}
\includegraphics[width=8cm]{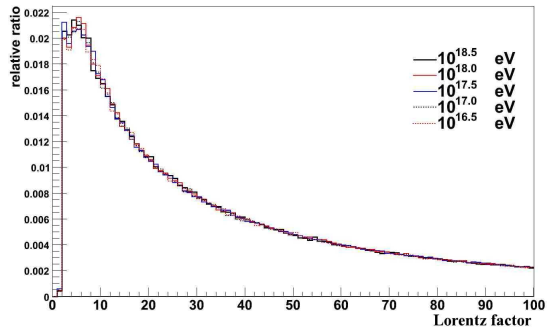}
\caption{Normalized energy distribution of the shower particles at the shower maximum for different shower energies. The particle energy is represented by its Lorentz factor. }
\label{gamma}
\end{center}
\end{figure}

\begin{figure}[htb]
\begin{center}
\includegraphics[width=8cm]{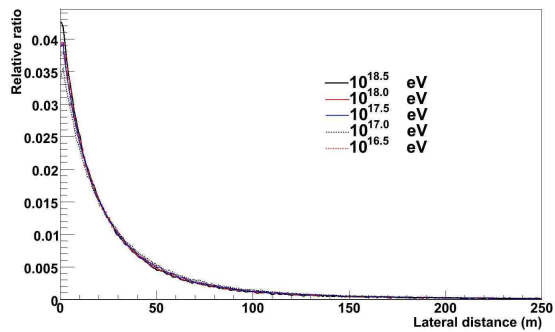}
\caption{Normalized lateral distribution of the shower particles at the shower maximum for different shower energies. The lateral distance is from the shower illuminated area to the observational site.}
\label{lateral}
\end{center}
\end{figure}

Fig. \ref{gamma} and \ref{lateral} show the energy and position distributions of the shower particles at shower maximum for different energies. Both are displayed in the unit normalized to the total number at the corresponding energies. 

\section{Radio Properties}

\begin{figure}[htb]
\begin{center}
\includegraphics[width=8cm]{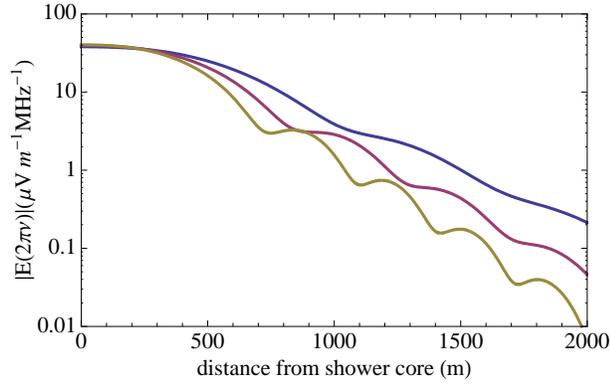}
\caption{Off-set dependence of $|E(R,2\pi\times\nu)|$ for the maximum of a $10^{17}{\rm eV}$ shower at the observation distance of $10{\rm km}$. Curves in blue, red and yellow represent signals in observing frequencies of $50{\rm MHz}$, $75{\rm MHz}$ and $100{\rm MHz}$, respectively.}
\label{interfere}
\end{center}
\end{figure}

\begin{figure}[htb]
\begin{center}
\includegraphics[width=8cm]{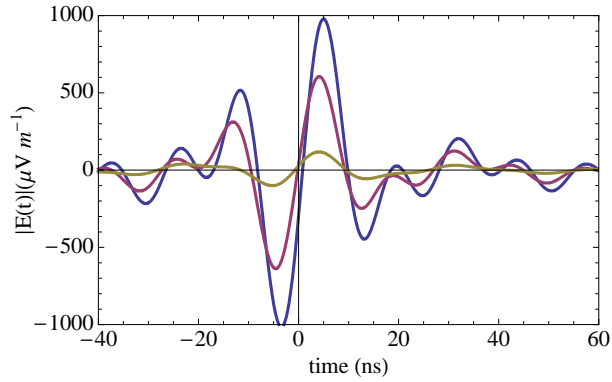}
\caption{Reconstructed pulses from emission of a $10^{17}{\rm eV}$ shower at the observation distance of $10{\rm km}$, using an idealized rectangular filter spanning $30-80{\rm MHz}$.  Curves in blue, red and yellow denote pulses measured at center, at lateral distances of $500{\rm m}$ and $1000{\rm m}$,  respectively.}
\label{pulse}
\end{center}
\end{figure}

Having determined the spatial structure and energy distribution of shower particles, we can calculate the emission from the shower maximum. Fig. \ref{interfere} depicts the expected electric field at different receiver locations. The interference pattern arises from the scale of the shower front. In Fig. \ref{pulse}, we calculate the pulse measured by the receiver with a given bandwidth. This plot indicates how large the separation between antennae can be for the current technology.

\section{Summary}

In this work, we investigate properties of the earth-skimming tau neutrino induced shower. The universal behavior of the shower particle allows a simple parametrization which will be helpful in future calculations of the geosynchrotron radiation. Our calculations also provide useful information for  the future experiments.

\acknowledgements

KCL and GLL are supported by the National Science Council (NSC 096-2811-M-009-024 and NSC 096-2112-M-009-023-MY3, respectively) of Taiwan, R.O.C.

\end{document}